\documentclass[11pt,twoside]{article}
\usepackage{asp2004}
\usepackage{psfig}
\usepackage{epsf}
\usepackage{graphics}
\usepackage{lscape}
\def\edcomment#1{\iffalse\marginpar{\raggedright\sl#1\/}\else\relax\fi}
\reversemarginpar

\begin{document}
\title{Chandra X-ray Observations of WZ~Sge in Superoutburst}
\author{Peter J.\  Wheatley$^1$ and Christopher W.\  Mauche$^2$}
\affil{$^1$Dept.\  of Physics and Astronomy, University of Leicester, Leicester, UK\\
$^2$Lawrence Livermore Nat'l Lab., 7000 East Ave., Livermore, CA, USA
}

\begin{abstract}
We present seven separate {\it Chandra} observations of the 2001 
superoutburst of WZ~Sge. The high-energy outburst was dominated by intense 
EUV emission lines, 
which we interpret as boundary layer emission scattered into 
our line of sight in an accretion disc wind. The direct boundary layer 
emission was hidden from view, presumably by the accretion disc. 
The optical {\em outburst orbital hump} was detected 
in the EUV, but the {\em common superhump} was not, indicating a 
geometric mechanism in the former and a dissipative mechanism in the latter. 
X-rays detected during outburst were not consistent with boundary layer 
emission and we argue that there must be a second source of X-rays in dwarf 
novae in outburst. 
\end{abstract}
\thispagestyle{plain}

\vspace{-0.1cm}
\section{Introduction}
WZ~Sge is a nearby, short period dwarf nova with an exceptionally long 
inter-outburst interval (20--30\,yr). 
These long intervals are 
probably
caused by the truncation of the 
(usually unstable) 
inner accretion disc by the magnetic field of the white dwarf
\citep{Warner96,Hameury97}. This picture is supported by the detection of 
a 28\,s modulation in the quiescent optical and X-ray emission, 
which is probably the spin period of the magnetised 
white dwarf \citep{Patterson98}.

The July--August 2001 superoutburst of WZ~Sge 
was the first since 1978, and 
was observed intensively
from the ground and 
from space 
\citep[e.g.,][]{Patterson02,Knigge02,Long03,Sion03}.
In this paper we present the X-ray and extreme-ultraviolet (EUV) observations 
made with the {\it Chandra} observatory. 
These data were originally reported by 
\citet{Wheatley01} and \citet{Kuulkers02}. 

\section{Observations}
\begin{figure}[!ht]
\plotfiddle{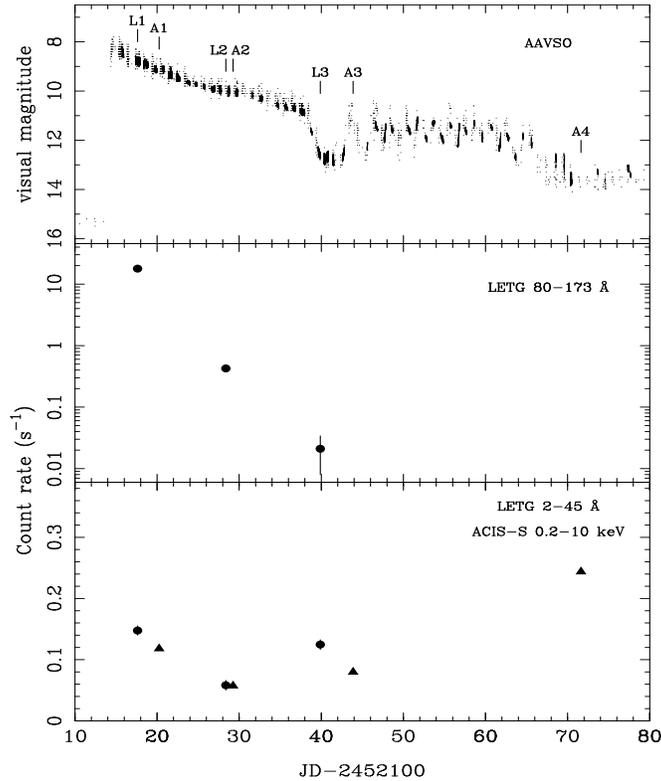}{9.2cm}{0}{60}{51}{-130}{-30}
\caption{
Optical, EUV, and X-ray light curves of 
the 2001 superoutburst of WZ~Sge. 
Top: the AAVSO visual light curve. 
Middle: the {\it Chandra\/} LETG 80--173\,\AA\  EUV light curve
plotted on the same dynamic range as the visual light curve. 
Bottom: the {\it Chandra\/} LETG 2--45\,\AA\  X-ray light curve (circles) and 
the ACIS-S 0.2--10\,keV X-ray light curve (triangles).
The ACIS-S count rates have been scaled by a factor 1/17 in order to align the 
A2 and L2 count rates. 
The times of the {\it Chandra\/} observations are 
labelled on the AAVSO panel, with L indicating a 
LETG observation and A indicating an ACIS-S observation.
\label{fig-all}
}
\end{figure}

WZ~Sge was observed seven times with {\it Chandra} during the 2001 
superoutburst. Three observations were made using the Low Energy Transmission 
Grating (LETG) and four with the Advanced CCD Imaging Spectrometer (ACIS-S). 
LETG observations provide high spectral resolution in the 
EUV and soft X-ray wavebands (5--170\,\AA). ACIS-S provides low-resolution 
observations in the hard and soft X-ray wavebands (1--50\,\AA). 
The times of all seven observations are indicated with respect to 
the optical outburst in the top panel of Fig.\,\ref{fig-all}. 
The LETG observations are identified with the labels L1--3 and the 
ACIS-S observations with A1--4. 

\section{Light curves}
The lower panels of Fig.\,\ref{fig-all} show the mean count rates 
from the seven {\it Chandra} observations. It can be seen that the 
EUV flux was very high in the early outburst
and then declined more quickly than the optical. 
The X-ray flux dropped gradually during the main outburst, recovered 
after the end of the main outburst (L3), but was suppressed
again during the first echo outburst (A3). The X-ray flux recovered again 
(by a larger factor) by the end of the echo outbursts (A4). 
This overall EUV and X-ray flux evolution is broadly consistent with that 
seen in other dwarf novae in outburst \citep[e.g.,][]{Wheatley03-ss}.

Light curves from the first 11\,d of the optical outburst were dominated by an 
{\em outburst orbital hump} \citep[OOH;][]{Patterson02}. This is believed to 
be due to the excitation of the 2:1 tidal resonance in the expanding accretion 
disc, and is seen only in WZ~Sge-type stars, where the mass 
ratio is sufficiently extreme to allow the disc to reach the resonant 
radius \citep{Osaki02,Kato02}. 
The later optical outburst is dominated by the {\em common superhump,} 
which is 
a feature of the superoutbursts in all dwarf novae, and is believed to be due 
to the excitation of the 3:1 tidal resonance. 

\begin{figure}[!t]
\plotfiddle{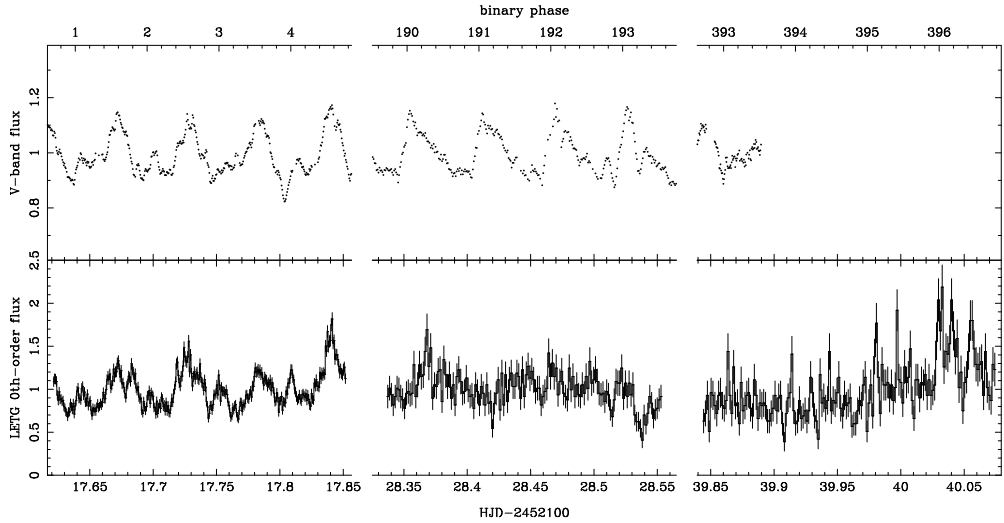}{5.75cm}{0}{69}{69}{-215}{-20}
\caption{
Optical and {\it Chandra\/} LETG light curves from observations
L1--3. Top: optical light curve in flux units $10^{-0.4(V-C)}$,
where $V$ ($C$) is WZ Sge's (the comparison star's) visual magnitude,
normalised to one over the interval of overlap with the LETG data. 
Bottom: LETG 0th-order light curve binned at 64 s (L1) and 128 s (L2 and L3) 
and normalised to one in each observation. 
Optical data were originally presented by 
\citet{Patterson02}.
\label{fig-llc}
}
\end{figure}

The light curves of the individual LETG observations, plotted in 
Fig.\,\ref{fig-llc}, show that 
the OOH is detected in the EUV, but that the common 
superhump is not. It can seen that the OOH modulations in the optical and EUV 
are both double peaked and in phase. 
These results suggest that the OOH is a geometric effect, 
with a two-arm spiral wave 
modifying our view of
the inner accretion disc and/or white dwarf. The absence of the common 
superhump in the EUV emission supports the view that this is a 
dissipative effect in the eccentric accretion disc.

\begin{figure}[!ht]
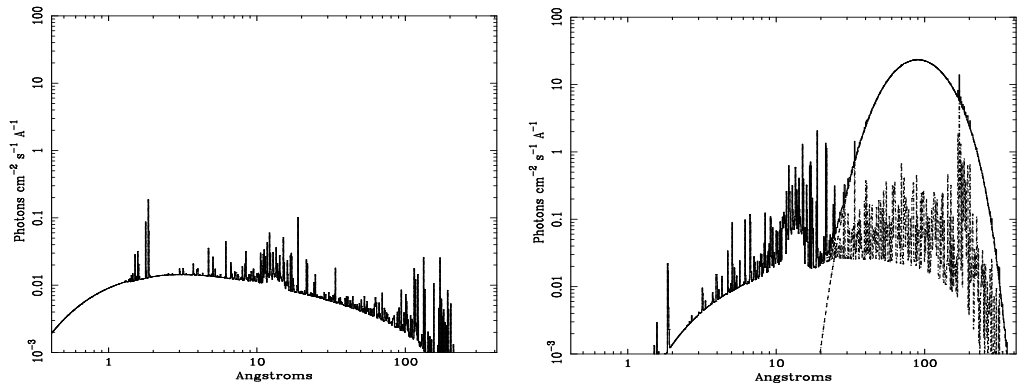

\plotfiddle{wheatley_f3a.ps}{3cm}{0}{74}{56}{-220}{-47}
\plotfiddle{wheatley_f3b.ps}{0cm}{0}{37}{28}{-15}{-18}
\caption{
Schematic X-ray spectra of dwarf novae in quiescence (left) and outburst 
(right).
\label{fig-schematic}
}
\end{figure}

\section{Spectra}
The high-energy spectra of dwarf novae change dramatically between quiescence 
and outburst. This is illustrated in Fig.\,\ref{fig-schematic}.
Quiescent X-ray spectra are normally fitted with multi-temperature 
optically thin thermal models, such as cooling flows, with 
temperatures in the range 1--10\,keV \citep[e.g.,][]{Mukai03}.
In outburst the X-ray emission is replaced with 
intense optically thick emission in the EUV band, with characteristic 
temperatures around 10\,eV \citep[e.g.,][]{Mauche04}, 
and an X-ray spectrum that is lower in temperature than
in quiescence \citep[e.g.,][]{Baskill04}.

\begin{figure}[!ht]
\plotfiddle{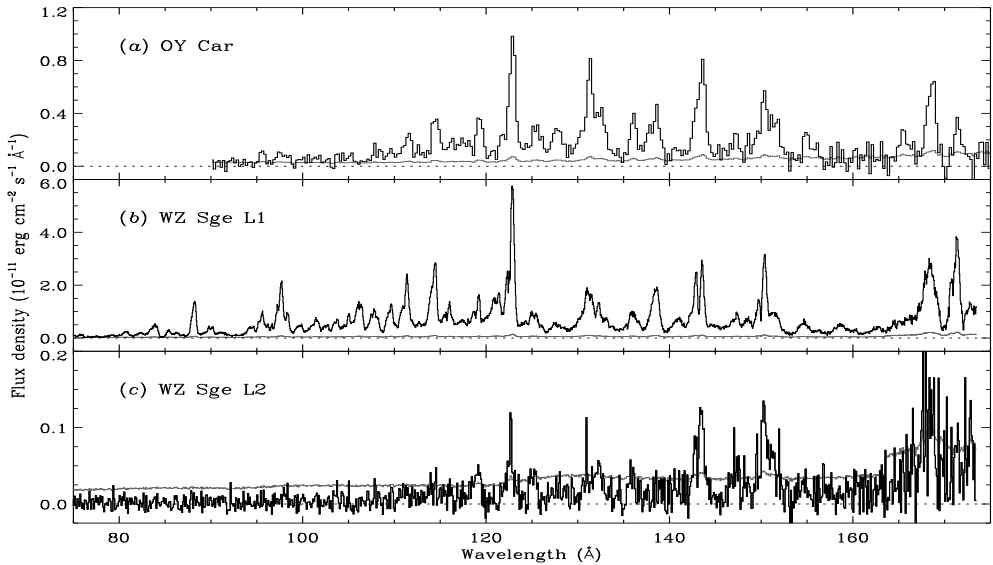}{6.75cm}{0}{80}{60}{-190}{-10}
\caption{
{\it EUVE\/} SW spectrum of OY Car in superoutburst ({\it a\/}) 
and {\it Chandra\/} LETG spectra of WZ Sge from observations L1 ({\it
b\/}) and L2 ({\it c\/}). Data are binned at 0.25, 0.05, and 0.1~\AA , 
respectively. Grey histograms are the associated $1\,\sigma $ error
vectors. {\it EUVE\/} data were originally presented by 
\citet{Mauche00}.
\label{fig-oy_l1_l2}
}
\end{figure}

The L1 and L2 EUV spectra of WZ~Sge in outburst are plotted in the 
lower panels of
Fig.\,\ref{fig-oy_l1_l2}.
Instead of the expected bright continuum these spectra are 
dominated by strong and broad emission lines. The only similar EUV spectrum 
of a dwarf nova is the 
{\it EUVE\/} spectrum of the eclipsing system OY~Car in superoutburst, 
which is reproduced in the top panel of Fig.\,\ref{fig-oy_l1_l2}. The 
similarity is striking.
\citet{Mauche00} 
argued that this spectrum
can be understood if the boundary layer of OY~Car is hidden from direct
view by the highly-inclined accretion disk, 
and if boundary layer radiation is scattered
into our line of sight by the accretion disk wind. 
This interpretation naturally explains the
salient features of the EUV spectra of OY~Car and WZ~Sge in outburst: 
the lines are identified
with resonance transitions of intermediate ionisation stages of cosmically
abundant elements, the lines are broad 
($\rm FWHM\approx 1\,\AA \sim 3000\,km\,s^{-1}$) 
because the wind velocity is high, and the continuum
is weak because the wind electron optical depth is low. 
The inclination of WZ~Sge is 
$77^\circ\pm2^\circ$ \citep{Spruit98} 
and so this interpretation requires that optically thick material extends 
at least $11^\circ$ out of the orbital plane. 

\begin{figure}[!t]
\plotfiddle{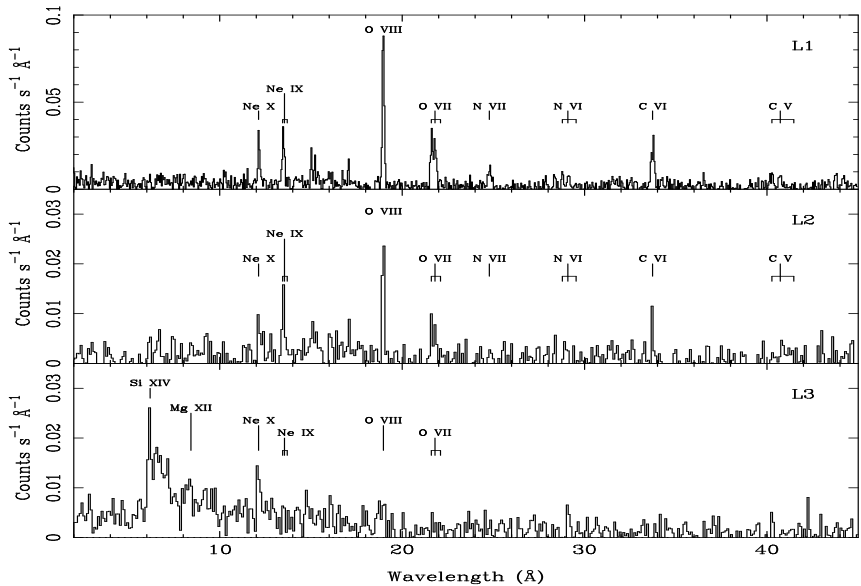}{6.75cm}{0}{60}{40}{-180}{-20}
\caption{
The short-wavelength LETG spectra from all three observations (L1--3).
The X-ray spectrum in the first two observations is dominated by emission 
lines from H-like and He-like species of C, N, O, Ne. 
These are labelled. Other lines include those of the Fe L-shell (15--18\,\AA).
The third spectrum also contains thermal lines, but the line ratios (e.g.,\   
Ne) show that the temperatures are much higher. 
\label{fig-l123_xray}
}
\end{figure}

The LETG X-ray spectra during outburst are plotted in 
Fig.\,\ref{fig-l123_xray}. The first two spectra are 
dominated by emission lines of 
H- and He-like species of C, N, O, and Ne. 
These spectra are therefore likely to represent thermal emission from 
a collisionally-excited optically thin plasma. 
The third LETG spectrum, taken late in the decline from the main outburst 
and before the first echo outburst, shows a blue continuum and strong H-like 
lines (note in particular the reversed H- and He-like line ratios of Ne).
This spectrum clearly originates from a plasma that is much hotter 
than that responsible for the L1 and L2 short-wavelength spectra.
Fitting this third spectrum, we find an acceptable fit with a 
single-temperature thermal plasma model with $kT=6.0\pm^{4.0}_{1.7}$\,keV and 
$N_{\rm H}=(1.4\pm0.5)\times10^{21}\rm\,cm^{-2}$. This temperature is 
typical of the boundary layers of dwarf novae in quiescence, 
while the high column density is presumably due to the remnants of the disc 
material that blocked our view of the boundary layer during outburst. 

The L1 and L2 spectra, in contrast, are not consistent with emission from the 
boundary layer. Spectral fitting shows that the emission measure rises to 
lower temperatures, whereas boundary layer emission in quiescence rises to high 
temperatures \citep[e.g.,][]{Mukai03}. More importantly, the L1 and L2 spectra 
are not consistent with the high column density measured in the L3 spectrum. 
This shows that the X-ray emission seen in outburst cannot pass through the 
material blocking our view to the boundary layer. 

Other possible sources of X-ray emission during outburst include scattering in 
the accretion disc wind (as we suggest for the EUV component). This has 
also been suggested to explain the lack of soft X-ray eclipses in high-state 
systems \citep[e.g.,][]{Pratt04}. 
However, unlike the EUV, there is no evidence for a strong continuum 
in soft X-rays
that could be scattered by the wind. The soft X-ray lines also 
arise from more highly ionised species than are believed to be present in the 
wind, and crucially,  the X-ray lines in WZ~Sge are too narrow to be due to 
scattering in the wind ($800\,\rm km\,s^{-1}$
compared with $3000\,\rm km\,s^{-1}$ in the EUV).

Alternative possibilities for the origin of soft X-rays during dwarf nova 
outbursts include internal shocks in the $3000\,\rm km\,s^{-1}$ wind, and 
solar-like coronal emission from the atmosphere of the accretion disc itself
($800\,\rm km\,s^{-1}$ corresponds to the Kepler velocity 
of the accretion disc at a radius of around $2\times10^{10}\,\rm cm$). 
Coronal emission might turn on in outburst as the disc becomes more 
magnetically active. 
In this case 
soft X-ray outburst emission might allow us to study 
the magneto-rotational instability in action.

\acknowledgements{We thank the {\it Chandra\/} director Harvey Tannenbaum and 
his deputy Fred Seward for their allocation of Director's Discretionary Time 
to this programme.
We also thank Joe Patterson 
for providing optical data, 
and we acknowledge with thanks the observations from the
AAVSO International Database contributed by observers worldwide and
used in this research.
PJW's contribution to this work was supported by PPARC rolling grants held at 
the University of Leicester. 
CWM's contribution to this work was performed under the auspices
of the U.S.\ Department of Energy by University of California Lawrence
Livermore National Laboratory under contract No. W-7405-Eng-48. }

\end{document}